\documentclass[9pt,conference]{IEEEtran}
\usepackage{dcase2026}

\usepackage{bm} 

\usepackage{dcase2026,amsmath,graphicx,url,times,booktabs, tabularx}

\title{Sampling Bias Compensation for Robust Evaluation of Audio Classification Systems with Partially Labeled Evaluation Datasets}

\name{Javier Naranjo-Alcazar$^{1}$,
      Annamaria Mesaros$^{2}$,
      Tuomas Virtanen$^{2}$,
      Pedro Zuccarello$^{1}$}
\address{$^{1}$Instituto Tecnologico de Informatica (ITI), Valencia, Spain  \\
$^{2}$Tampere University, Tampere, Finland\\
}

\begin{document}

\maketitle

\begin{abstract}
The performance of acoustic machine learning systems is commonly evaluated using fully annotated test sets. In real-world deployments, however, exhaustively labeling large volumes of continuously collected audio data is often infeasible. Consequently, performance assessment typically relies on a small labeled subset of the available data, introducing a sampling bias that can severely distort evaluation metrics. This paper studies methods for compensating the bias in evaluation-labeled subsets under strict annotation-budget constraints. We study whether importance weighting techniques can mitigate this discrepancy by compensating for the selection bias. Specifically, we implement and compare three density-ratio estimation methods: kernel density estimation (KDE), logistic regression, and k-nearest neighbors (kNN), utilizing feature-space representations of the deployed audio. To emulate realistic deployment scenarios, the labeled subsets are generated using five distinct sampling strategies based on active learning techniques. Experiments conducted on an audio scene classification (ASC) benchmark demonstrate that importance weighting consistently yields more realistic accuracy estimates, significantly reducing the gap between subset-based metrics and the true evaluation performance. 
\end{abstract}


\section{Introduction}
\label{sec:intro}



The performance of machine listening systems is traditionally assessed using fully annotated benchmark datasets under controlled experimental conditions \cite{fonseca2021fsd50k, piczak2015esc, politis2020dataset}. While this offline setting is appropriate for algorithm development, it rarely reflects the operational reality of real-world deployments. Once a model is trained on development data and deployed into production, it typically operates on continuous audio streams to perform ongoing inference \cite{Naranjo:2025:AES}. In practical monitoring applications, evaluating operational performance relies on analyzing a collected sample of this deployment traffic. However, because exhaustively annotating every operational recording is logistically and economically infeasible, performance assessment must rely on a small, selected subset of the ongoing stream. Consequently, tracking the model's long-term behavior depends entirely on evaluation metrics computed over this limited labeled snapshot, even though its representativeness with respect to the true deployment distribution remains uncertain.


A representative example of this operational challenge can be found in large-scale environmental audio monitoring initiatives such as Soroll-IA \cite{naranjo2025data}, where extensive recording campaigns continuously generate substantially more audio than can realistically be annotated. In these real-time streaming scenarios, annotation efforts are usually guided by active learning (AL) strategies or other sample selection procedures designed to identify segments that maximize annotation value \cite{shuyang2020active}. Consequently, the final evaluation subset is not obtained through random or exhaustive labeling, but rather by selecting a limited number of segments from a massive, continuously growing pool of collected audio. Moreover, the segments selected for annotation are typically chosen according to predefined criteria, such as maximizing diversity \cite{shuyang2020active, zhang2025hybrid}, prioritizing uncertain samples \cite{yang2016active, nguyen2022measure}, or increasing the representation of rare acoustic conditions \cite{jung2024active}. While these strategies are beneficial for building effective training datasets, they may produce a labeled subset whose distribution differs from that of the complete collection of recorded audio. This mismatch introduces selection bias, leading to evaluation metrics that are not fully representative of the performance that would be obtained if all collected segments were labeled.

This operational reality raises a critical question that has received little attention in the computer audition literature: how representative are the performance metrics computed on this selected subset with respect to the true performance over the entire stream of deployed data? When only a biased subset of the evaluation data is labeled, the estimated performance is inevitably affected by the sampling bias introduced by the selection procedure itself. As a result, the reported metric may fail to accurately reflect the actual behavior of the deployed system over its full operational distribution, potentially masking model degradation or yielding heavily optimistic performance estimates.

This challenge is particularly critical for model monitoring, maintenance decisions, and system verification in production, where obtaining reliable performance estimates under realistic annotation budgets remains an open problem. To address this, we investigate the problem of performance estimation from partially annotated evaluation datasets in acoustic scene classification (ASC) \cite{ding2024acoustic, chen2025improving, liang2026tf}. Rather than treating all labeled samples equally, we exploit the feature-space representations of the pool of unlabeled data to correct the selection bias. Specifically, we estimate density ratios between the full deployment distribution and the labeled subset to compute importance weights, which are then used to calculate weighted evaluation metrics.

In this work, we implement and evaluate three distinct density-ratio estimation methods for this task: kernel density estimation (KDE) \cite{nair2019covariate, sugiyama2007covariate}, logistic regression \cite{nair2019covariate}, and k-nearest neighbors (kNN) \cite{kanamori2009least}. To rigorously test these methods, we simulate realistic deployment conditions on an ASC benchmark by creating evaluation subsets under five different sampling strategies based on active learning techniques, alongside standard random sampling. Our objective is not to improve classification accuracy itself, but rather to provide a reliable framework for system monitoring when exhaustive annotation is impossible \cite{zadrozny2004learning,learning2009dataset}.


Our results demonstrate that importance weighting consistently yields realistic accuracy estimates, significantly reducing the gap between subset-based metrics and true performance across various budgets. To our knowledge, this is the first work in computer audition to explicitly employ importance weighting to correct selection bias for deployment evaluation.

\section{Method}
\label{sec:method}

\subsection{Overview}
\label{subsec:overview}

Let $\mathcal{D}$ denote the complete deployment distribution of paired audio samples and their respective labels. To clearly contextualize this scenario, we distinctly separate the development stage—where the model is traditionally trained, tuned, and validated using localized train, validation, and test partitions—from the operational deployment stage. In this latter phase, the frozen model is released into production to process streams of unseen, unannotated real-world audio. In practice, due to strict budget constraints, only a small fraction of these continuous segments can be manually annotated. Moving forward, we refer to this complete pool of operational data interchangeably as the deployment dataset or the evaluation dataset, and we denote the small, subsequently labeled subset selected from it as $\mathcal{L}=\{(x_i,y_i)\}_{i=1}^{n},$ where $x_i$ represents the audio segment and $y_i$ is its associated label. The remaining segments $\mathcal{U}=\{{x_j}\}_{j=1}^{m}$ are assumed to be unlabeled.

The objective of this work is not to optimize the accuracy of the underlying audio classification model itself, but rather to obtain a more reliable and realistic estimate of its deployment performance by leveraging the limited labeled subset $\mathcal{L}$ alongside the structural distribution information contained in the unlabeled pool $\mathcal{U}$. Ideally, the true performance metric $M_{\mathrm{full}}$ over the entire deployment distribution $\mathcal{D}$ is defined as the expected value:

\begin{equation}
M_{\mathrm{full}} = \mathbb{E}_{(x,y) \sim \mathcal{D}} [m(x, y)],
\end{equation}

\noindent where $m(x, y)$ represents the scoring function for a given pair. In practice, however, $M_{\mathrm{full}}$ cannot be directly computed because the complete deployment dataset is unlabeled. Instead, a common workaround is to calculate the standard empirical estimate solely from the available labeled subset $\mathcal{L}$, which is defined as:

\begin{equation}
\hat{M}_{\mathrm{subset}} = \frac{1}{n} \sum_{i=1}^{n} m(x_i, y_i).
\end{equation}

\noindent However, if $\mathcal{L}$ is constructed via a biased sampling strategy, $\hat{M}_{\mathrm{subset}}$ becomes a biased estimator of $M_{\mathrm{full}}$. To compensate for this selection bias, importance weighting is employed to compute a corrected weighted metric estimate:

\begin{equation}
\hat{M}_{\mathrm{IW}} = \frac{\sum_{i=1}^{n} w_i m(x_i, y_i)}{\sum_{i=1}^{n} w_i},
\end{equation}
where $w_i$ is the importance weight that quantifies how representative sample $x_i$ is with respect to the deployment distribution.

In this work, we operationalize this generic framework focusing on the classification micro-accuracy. Thus, by defining the scoring function through the indicator function $m(x_i, y_i) = \mathbb{I}(\hat{y}_i = y_i)$—where $\hat{y}_i$ is the prediction yielded by the model—the unweighted accuracy estimate $\hat{A}_{\mathrm{subset}}$ and the importance-weighted micro-accuracy estimate $\hat{A}_{\mathrm{IW}}$ are formalized respectively as:
\begin{equation}
\hat{A}_{\mathrm{subset}} = \frac{1}{n} \sum_{i=1}^{n} \mathbb{I}(\hat{y}_i = y_i),
\end{equation}
\begin{equation}
\hat{A}_{\mathrm{IW}} = \frac{\sum_{i=1}^{n} w_i \mathbb{I}(\hat{y}_i = y_i)}{\sum_{i=1}^{n} w_i}.
\label{eq:aiw}
\end{equation}

Theoretically, the ideal importance weight for a given sample $x_i$ is defined as the density ratio between the full deployment distribution and the selection-biased distribution:
\begin{equation}
w(x_{i}) = \frac{p_{\mathcal{D}}(x_{i})}{p_{\mathcal{L}}(x_{i})},
\end{equation}
where $p_{\mathcal{D}}(x)$ represents the probability density function of the deployment distribution and $p_{\mathcal{L}}(x)$ represents the density function induced by the labeled subset selection. In practice, these densities are unknown, and the operational weights $w_i$ must be estimated using the unlabeled pool $\mathcal{U}$ and the labeled subset $\mathcal{L}$. To avoid introducing additional feature engineering steps, all density-ratio estimation strategies investigated in this work operate directly on the embedding vectors extracted from the baseline classification model.

\subsection{Kernel Density Estimation}

The first approach estimates the density ratio through kernel density estimation (KDE). Direct density estimation in the original embedding space was found to be unstable due to the curse of dimensionality, which causes KDE estimates to collapse in high-dimensional settings. To guarantee numerical stability while preserving the core structural patterns of the space, the embeddings are first standardized using the deployment data mean and variance, and subsequently projected onto their first eight principal components via principal component analysis (PCA) with 8 components. Two independent KDE models are then fitted to this reduced representation, modeling the deployment distribution $\hat{p}_{\mathcal{D}}(z)$ and the distribution induced by the labeled subset $\hat{p}_{\mathcal{L}}(z)$, respectively. The importance weight $w_i$ associated with a specific labeled sample embedding $z_i$ is computed in the logarithmic domain to prevent numerical underflow issues:
\begin{equation}
\log w_i = \log \hat{p}{\mathcal{D}}(z_i) - \log \hat{p}{\mathcal{L}}(z_i).
\end{equation}Finally, the exponential weights are restored, and a normalization step is applied across the labeled subset to enforce a unit mean constraint, preventing scale mismatches in the subsequent metric computation:
\begin{equation}
w_i \leftarrow \frac{w_i}{\frac{1}{n} \sum_{k=1}^{n} w_k}.
\label{eq:normalize}
\end{equation}

\subsection{Logistic Regression Density Ratio Estimation}

The second approach reformulates density-ratio estimation as a binary classification problem. Samples from the deployment pool $\mathcal{D}$ and the labeled subset $\mathcal{L}$ are assigned domain labels $d = 1$ and $d = 0$, respectively. A logistic regression classifier is then trained on the combined sets of embedding vectors $z$ extracted from the audio classification model. Given an embedding vector $z_i$ corresponding to a sample from the labeled subset $\mathcal{L}$, the classifier estimates its posterior probability of belonging to the deployment distribution, denoted as $P(d=1|z_i)$. Under the probabilistic density-ratio framework, and noting that $P(d=0|z_i) = 1 - P(d=1|z_i)$, the unnormalized importance weight $w_i$ for each labeled sample is directly expressed as:
\begin{equation}
w_i = \frac{P(d=1|z_i)}{P(d=0|z_i)} = \frac{P(d=1|z_i)}{1 - P(d=1|z_i)}.
\end{equation}
By bypassing explicit density estimation, this discriminative formulation is inherently more robust than generative approaches like KDE when operating in high-dimensional embedding spaces. Finally, the extracted weights are normalized as per Eq.~\ref{eq:normalize}.

\subsection{k-Nearest Neighbour Density Ratio Estimation}

The third approach estimates the density ratio using local neighborhood statistics, bypassing explicit global density modeling. Given an embedding vector $z_i$ corresponding to a sample from the labeled subset $\mathcal{L}$, its local density $\rho(z_i)$ is assumed to be inversely proportional to its average distance to its $k$-nearest neighbors, denoted as $\bar{d}_{k}(z_i)$:

\begin{equation}\rho(z_i) \propto \frac{1}{\bar{d}_{k}(z_i)}.
\end{equation}

By estimating this local density independently with respect to the deployment pool ($\rho_{\mathcal{D}}(z_i)$) and the labeled subset ($\rho_{\mathcal{L}}(z_i)$), the sample importance weight $w_i$ directly simplifies to the ratio of their respective average neighbor distances:

\begin{equation}
w_i = \frac{\rho_{\mathcal{D}}(z_i)}{\rho_{\mathcal{L}}(z_i)} = \frac{\bar{d}{k}^{\mathcal{L}}(z_i)}{\bar{d}{k}^{\mathcal{D}}(z_i)},
\end{equation}

\noindent where $\bar{d}_{k}^{\mathcal{L}}(z_i)$ is the average distance from $z_i$ to its $k$-nearest neighbors within the labeled subset $\mathcal{L}$, and $\bar{d}_{k}^{\mathcal{D}}(z_i)$ is the average distance to its $k$-nearest neighbors within the deployment pool $\mathcal{D}$. Finally, the extracted weights are normalized as Eq.~\ref{eq:normalize}.

By relying strictly on local geometric distributions, this non-parametric formulation adapts better to irregular, multi-modal structures in the embedding space than generative kernel methods.

\section{Experimental Details}\label{sec:experiment}

\subsection{Dataset}

Experiments were conducted using the baseline benchmark from the DCASE 2017 Acoustic Scene Classification challenge \cite{DCASE2017challenge}. This dataset consists of 10-second binaural audio segments captured across 15 distinct acoustic scenes, spanning various transportation, indoor, and outdoor environments. Following the official evaluation protocol, the standard evaluation partition—comprising 1,620 samples with a balanced distribution of 108 segments per class—was treated as our complete deployment dataset. To simulate realistic operational conditions under strict annotation-budget constraints, we assume that only a small fraction of this partition is manually annotated to form the labeled subset $\mathcal{L}$, while the remaining data constitutes the unlabeled pool $\mathcal{U}$. The performance metrics computed over the entire, fully annotated evaluation partition serve as the ground-truth reference target that our importance-weighting methods aim to recover.

\subsection{Baseline Model}


The baseline classification model is a standard Convolutional Neural Network (CNN) adopted from the DCASE 2018 ASC challenge baseline~\cite{Mesaros2018_DCASE}. The network processes 10-second audio segments represented as $40 \times 500$ log-mel spectrograms (40 mel bands, 40 ms windows, 50\% overlap). It consists of two convolutional blocks followed by a fully connected classifier; complete structural specifications and training hyper-parameters can be found in~\cite{Mesaros2018_DCASE}. For the scope of this work, the model is frozen after development training to serve as a static feature extractor, with embeddings extracted from its penultimate dense layer (100 units).

\subsection{Evaluation Subset Construction}







To simulate realistic operational constraints, we evaluate the performance estimation methods across multiple labeled subset sizes. Specifically, the number of annotated samples $n$ (where $n = |\mathcal{L}|$) is varied systematically:

\[
n \in
\begin{aligned}
\{50,100,150,200,300,400,500,600\} \\
\cup \{800,900,1100,1250,1400,1550\}
\end{aligned}
\]

For each target size $n$, the designated number of segments is selected from the complete evaluation partition to form the labeled subset $\mathcal{L}$, while the remaining $1620 - n$ samples constitute the unlabeled pool $\mathcal{U}$. To account for the stochasticity inherent to sample selection—particularly in random sampling and diversity-based active learning strategies where initialization variations alter the final subset composition—all experiments are repeated 500 times using different random seeds. For every subset size and trial, the unweighted accuracy $\hat{A}_{\mathrm{subset}}$ and its corresponding importance-weighted estimate $\hat{A}_{\mathrm{IW}}$ are computed independently to assess statistical robustness.


\subsection{Subset Selection Strategies}

Several subset selection strategies were investigated to emulate different annotation procedures.

\begin{itemize}
\item \textbf{Random Sampling}
Samples are selected uniformly at random from the available pool. This strategy serves as a simple baseline and is frequently used to evaluate the effectiveness of more sophisticated selection approaches \cite{settles2009active}. This strategy provides an unbiased estimate of the density.

\item \textbf{Uncertainty Sampling}
Samples are selected according to the uncertainty of a classifier, prioritizing instances for which the model is least confident. In this work, uncertainty is estimated from the class posterior probabilities produced by the baseline acoustic scene classifier. Uncertainty sampling is one of the most widely adopted active learning strategies due to its simplicity and effectiveness \cite{settles2009active, shishkin2021active, meire2023active}.

\item \textbf{Farthest Traversal}
Samples are selected to maximize their minimum distance to the already selected set, promoting diversity and coverage of the feature space. This strategy is closely related to core-set selection methods and has been widely adopted in active learning and dataset summarization \cite{shuyang2020active}.

\item \textbf{K-Medoids Clustering}
The embedding space is partitioned into \(k\) clusters and the medoid of each cluster is selected. Unlike centroid-based methods, medoids correspond to actual observations and provide a representative subset of the underlying data distribution \cite{shuyang2017active}.

\item \textbf{Density-Weighted Farthest Traversal}
This method combines diversity and representativeness by weighting the Farthest Traversal criterion with a local density estimate. Consequently, samples are selected not only because they are distant from previously chosen examples but also because they belong to densely populated regions of the feature space, reducing the risk of selecting isolated outliers \cite{settles2008analysis}.
\end{itemize}

These strategies were chosen because they represent commonly used principles in active learning and dataset curation, namely random selection, diversity maximization, cluster representativeness, density awareness, and uncertainty-driven exploration. In conclusion, the final importance-weighted accuracy estimates $\hat{A}_{\mathrm{IW}}$ are computed via Eq.~\ref{eq:aiw}. For each experimental configuration, the weights $w_i$ are estimated independently using the density-ratio techniques detailed in Section~\ref{sec:method} across the localized subsets $\mathcal{L}$ extracted by the sampling mechanisms presented above.

\begin{figure*}
    \centering
    \includegraphics[width=1\linewidth]{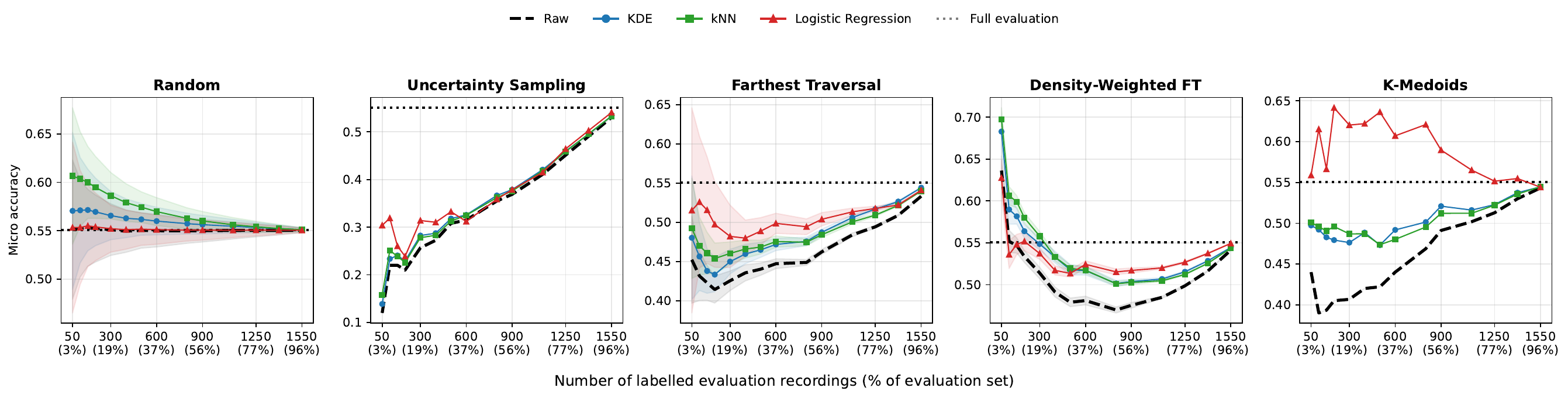}
    \caption{Micro-accuracy under partial labeling across different selection strategies. Dotted line: reference performance ($A_{\mathrm{full}}$). Dashed line: unweighted accuracy ($\hat{A}_{\mathrm{subset}}$). Colored curves: importance-weighted estimates ($\hat{A}_{\mathrm{IW}}$) via KDE, kNN, and Logistic Regression. Shaded regions indicate $\pm1$ SD over 500 independent runs.}
    \label{fig:results}
\end{figure*}

\section{Results}
\label{sec:results}

Fig.~\ref{fig:results} presents the micro-accuracy estimates obtained across varying labeled subset sizes $n$. The horizontal dotted black line represents the true performance computed over the complete evaluation dataset, serving as the reference target ($A_{\mathrm{full}}$). The dashed black line tracks the unweighted accuracy $\hat{A}_{\mathrm{subset}}$ calculated directly from the biased labeled subset without correction. Finally, the colored curves depict the importance-weighted accuracy estimates $\hat{A}_{\mathrm{IW}}$ yielded by the three proposed density-ratio estimation methods: KDE, Logistic Regression, and kNN. Based on these results, several key observations can be made.

First, the accuracy estimated from the labeled subset is strongly affected by the sampling strategy used to construct it. Since the DCASE 2017 evaluation partition is perfectly balanced and does not exhibit intrinsic data distribution skewness, the average unweighted accuracy $\hat{A}_{\mathrm{subset}}$ under Random Sampling converges close to the full-evaluation baseline performance when averaged over 500 independent runs. However, this statistical convergence masks a critically high variance across individual trials, rendering Random Sampling highly unreliable in practice. Furthermore, in realistic production pipelines characterized by massive streams of unannotated data, Random Sampling is highly impractical, as it frequently selects redundant, low-information samples. Conversely, structured active learning strategies—such as uncertainty sampling, Farthest Traversal (FT), Density-weighted FT, and K-Medoids clustering—are specifically designed to maximize informational utility. Yet, as observed in Fig.~\ref{fig:results}, these targeted selection mechanisms inherently introduce substantial estimation biases. This result highlights that, in restricted label regimes, the reported performance of a model may depend more on the sample selection mechanism itself than on the actual operational capabilities of the classifier.

A prominent finding is the active response of the Logistic Regression-based density ratio estimator ($\hat{A}_{\mathrm{IW}}$ with Logistic Regression, depicted in red), which stands out as the most assertive weighting method, though its behavior is highly dependent on the underlying sampling mechanism. Across the diversity-driven strategies based on Farthest Traversal (FT and Density-Weighted FT), the Logistic Regression weights systematically provide the most effective correction, successfully pushing the underestimated unweighted accuracy ($\hat{A}_{\mathrm{subset}}$) upwards and aligning it closely with the true reference baseline ($A_{\mathrm{full}}$) in the low-label regime ($n \leq 300$). However, this aggressive corrective behavior leads to a noticeable overestimation in the case of K-Medoids clustering, where the Logistic Regression estimate yields values substantially higher than $A_{\mathrm{full}}$ before asymptotically returning to the baseline at larger subset sizes. This overestimation, contrasted with its complete lack of correction under Uncertainty Sampling (where it remains strictly bounded to $\hat{A}_{\mathrm{subset}}$), suggests that while the discriminative nature of Logistic Regression is highly capable of driving variance in high-dimensional spaces, its accuracy is bounded by how the sampling strategy populates the geometric boundaries of the latent space.

Conversely, the KDE-based (blue) and kNN-based (green) weighting approaches yield more conservative and localized modifications, though their effectiveness varies significantly across sampling strategies. In the Random Sampling scenario, both non-parametric methods tend to slightly overestimate the true baseline performance in the low-budget regime. Under the Farthest Traversal variants (FT and Density-Weighted FT), they successfully provide a positive correction to the underestimated unweighted baseline ($\hat{A}_{\mathrm{subset}}$), albeit remaining visibly below the optimal recovery achieved by Logistic Regression. Interestingly, their most remarkable performance is observed under K-Medoids clustering. In this specific setting, while Logistic Regression suffers from a severe overestimation, both KDE and kNN demonstrate an optimal, well-bounded corrective behavior, tracking the true reference performance ($A_{\mathrm{full}}$) closely across almost all subset sizes. This localized success suggests that explicit, non-parametric density estimation in the latent space is highly adept at compensating for smooth, representative geometric partitions like those generated by K-Medoids, even if it struggles to match the dynamic range of discriminative models in more aggressively biased configurations.


Finally, as the labeling budget $n$ increases toward the full size of the evaluation set ($n = 1550$), a generalized convergence is observed across all subplots. As the labeled subset $\mathcal{L}$ becomes asymptotically representative of the entire deployment distribution, the unweighted accuracy $\hat{A}_{\mathrm{subset}}$ naturally aligns with $A_{\mathrm{full}}$. Consequently, the density ratios flatten towards unity, causing the corrective adjustments of all three importance-weighting methods ($\hat{A}_{\mathrm{IW}}$) to gracefully taper off.

\section{Conclusion}
\label{sec:conclusion}

This paper investigated performance estimation under partial evaluation labeling, a critical challenge in real-world acoustic monitoring deployments where manual annotations are severely budget-constrained. Using the Acoustic Scene Classification task from the DCASE 2017 challenge as a benchmark, we evaluated the capacity of three density-ratio estimation methods—KDE, kNN, and Logistic Regression—to correct the estimation bias introduced by various sample selection strategies and recover the true deployment performance ($A_{\mathrm{full}}$).

Our experimental results demonstrate that the unweighted accuracy $\hat{A}_{\mathrm{subset}}$ computed from a limited labeled subset $\mathcal{L}$ can deviate substantially from $A_{\mathrm{full}}$, driven more by the geometry of the selection mechanism than by the classifier's actual capabilities. In these biased regimes, importance weighting ($\hat{A}_{\mathrm{IW}}$) proved effective in reducing estimation errors using only the available unlabeled pool $\mathcal{U}$. 

The corrective behavior, however, is highly dependent on the interplay between the weighting and sampling strategies. The discriminative Logistic Regression approach provided the most dynamic and robust recovery across Farthest Traversal variants, although it exhibited a noticeable overestimation under K-Medoids clustering. Conversely, explicit non-parametric estimators (KDE and kNN) demonstrated a more conservative yet optimal, well-bounded alignment precisely under K-Medoids. Finally, we identified a hard operational boundary in Uncertainty Sampling, where the severe informational bottleneck prevents all density-ratio methods from mitigating the bias. In conclusion, while scarce evaluation labels demand caution, leveraging unlabeled deployment data via importance weighting substantially improves metric reliability without additional manual annotations.






\bibliographystyle{IEEEtran}
\bibliography{refs}

@string{icassp = "Proc. ICASSP"}

@string{dcase = "Proc. DCASE"}

@inproceedings{DCASE2017challenge,
    Author = "Mesaros, A. and Heittola, T. and Diment, A. and Elizalde, B. and Shah, A. and Vincent, E. and Raj, B. and Virtanen, T.",
    title = "{DCASE} 2017 Challenge Setup: Tasks, Datasets and Baseline System",
    booktitle = "Proceedings of the Detection and Classification of Acoustic Scenes and Events 2017 Workshop (DCASE2017)",
    year = "2017",
    month = "November",
    pages = "85--92"
}

@techreport{settles2009active,
  author       = {Burr Settles},
  title        = {Active Learning Literature Survey},
  institution  = {University of Wisconsin--Madison},
  type         = {Computer Sciences Technical Report},
  number       = {1648},
  year         = {2009},
  url          = {https://research.cs.wisc.edu/techreports/2009/TR1648.pdf}
}

@inproceedings{shuyang2017active,
  title={Active learning for sound event classification by clustering unlabeled data},
  author={Zhao, Shuyang and Heittola, Toni and Virtanen, Tuomas},
  booktitle={2017 IEEE International Conference on Acoustics, Speech and Signal Processing (ICASSP)},
  pages={751--755},
  year={2017},
  organization={IEEE}
}

@article{shuyang2020active,
  title={Active learning for sound event detection},
  author={Zhao, Shuyang and Heittola, Toni and Virtanen, Tuomas},
  journal={IEEE/ACM Transactions on Audio, Speech, and Language Processing},
  volume={28},
  pages={2895--2905},
  year={2020},
  publisher={IEEE}
}

@inproceedings{settles2008analysis,
  title={An analysis of active learning strategies for sequence labeling tasks},
  author={Settles, Burr and Craven, Mark},
  booktitle={Proceedings of the 2008 Conference on Empirical Methods in Natural Language Processing},
  pages={1070--1079},
  year={2008}
}

@inproceedings{shishkin2021active,
  title={Active learning for sound event classification using Monte-Carlo dropout and PANN embeddings},
  author={Shishkin, Stepan and Hollosi, Danilo and Doclo, Simon and Goetze, Stefan},
  booktitle={Proceedings of the 6th Workshop on Detection and Classication of Acoustic Scenes and Events (DCASE 2021)},
  pages={150--154},
  year={2021},
  organization={DCASE}
}

@inproceedings{meire2023active,
  title={Active learning in sound-based bearing fault detection},
  author={Meire, Maarten and Zegers, Jeroen and Karsmakers, Peter},
  booktitle={Proceedings of the 8th Detection and Classification of Acoustic Scenes and Events 2023 Workshop (DCASE2023), Tampere, Finland},
  pages={111--115},
  year={2023}
}

@inproceedings{piczak2015esc,
  title={ESC: Dataset for environmental sound classification},
  author={Piczak, Karol J},
  booktitle={Proceedings of the 23rd ACM International Conference on Multimedia},
  pages={1015--1018},
  year={2015}
}

@article{fonseca2021fsd50k,
  title={Fsd50k: an open dataset of human-labeled sound events},
  author={Fonseca, Eduardo and Favory, Xavier and Pons, Jordi and Font, Frederic and Serra, Xavier},
  journal={IEEE/ACM Transactions on Audio, Speech, and Language Processing},
  volume={30},
  pages={829--852},
  year={2021},
  publisher={IEEE}
}

@inproceedings{politis2020dataset,
    author = "Politis, Archontis and Adavanne, Sharath and Virtanen, Tuomas",
    title = "A Dataset of Reverberant Spatial Sound Scenes with Moving Sources for Sound Event Localization and Detection",
    booktitle = "Proceedings of the Detection and Classification of Acoustic Scenes and Events 2020 Workshop (DCASE2020)",
    address = "Tokyo, Japan",
    month = "November",
    year = "2020",
    pages = "165--169"
}

@inproceedings{Naranjo:2025:AES,
  author    = {J. Naranjo-Alcazar and J. Grau-Haro and R. Ribes-Serrano and P. Zuccarello},
  title     = {Real-Time Audio Monitoring Pipeline with Edge Inference and IoT Supervision via ThingsBoard},
  booktitle = {2025 AES International Conference on Artificial Intelligence and Machine Learning for Audio},
  publisher = {Audio Engineering Society},
  month     = {september},
  year      = {2025},
  address   = {London, UK},
  pages = "339--340",
  url       = {https://aes.org/wp-content/uploads/2026/01/AES-AIMLA-proceedings-final.pdf},
  note      = {Late Breaking Demo Paper}
}

@inproceedings{naranjo2025data,
  title={A Data-Centric Framework for Machine Listening Projects: Addressing Large-Scale Data Acquisition and Labeling through Active Learning},
  author={Naranjo-Alcazar, Javier and Grau-Haro, Jordi and Ribes-Serrano, Ruben and Zuccarello, Pedro},
  booktitle={Future of Information and Communication Conference},
  pages={647--659},
  year={2025},
  organization={Springer}
}

@article{ding2024acoustic,
  title={Acoustic scene classification: A comprehensive survey},
  author={Ding, Biyun and Zhang, Tao and Wang, Chao and Liu, Ganjun and Liang, Jinhua and Hu, Ruimin and Wu, Yulin and Guo, Difei},
  journal={Expert Systems with Applications},
  volume={238},
  pages={121902},
  year={2024},
  publisher={Elsevier}
}

@inproceedings{liang2026tf,
  title={TF-AttNet: An Efficient Time-Frequency Structure Modeling For Low-Complexity Acoustic Scene Classification},
  author={Liang, Yun and Luo, Tang and Chen, Zhichao and Zhong, Cunkun},
  booktitle={International Conference on Multimedia Modeling},
  pages={33--45},
  year={2026},
  organization={Springer}
}

@inproceedings{chen2025improving,
  title={Improving acoustic scene classification in low-resource conditions},
  author={Chen, Zhi and Shao, Yun-Fei and Ma, Yong and Wei, Mingsheng and Zhang, Le and Zhang, Wei-Qiang},
  booktitle={ICASSP 2025-2025 IEEE International Conference on Acoustics, Speech and Signal Processing (ICASSP)},
  pages={1--5},
  year={2025},
  organization={IEEE}
}

@inproceedings{nair2019covariate,
  title={Covariate shift: A review and analysis on classifiers},
  author={Nair, Nimisha G and Satpathy, Pallavi and Christopher, Jabez and others},
  booktitle={2019 Global Conference for Advancement in Technology (GCAT)},
  pages={1--6},
  year={2019},
  organization={IEEE}
}

@article{sugiyama2007covariate,
  title={Covariate shift adaptation by importance weighted cross validation.},
  author={Sugiyama, Masashi and Krauledat, Matthias and M{\"u}ller, Klaus-Robert},
  journal={Journal of Machine Learning Research},
  volume={8},
  number={5},
  year={2007}
}

@article{kanamori2009least,
  title={A least-squares approach to direct importance estimation},
  author={Kanamori, Takafumi and Hido, Shohei and Sugiyama, Masashi},
  journal={The Journal of Machine Learning Research},
  volume={10},
  pages={1391--1445},
  year={2009},
  publisher={JMLR. org}
}

@inproceedings{zadrozny2004learning,
  title={Learning and evaluating classifiers under sample selection bias},
  author={Zadrozny, Bianca},
  booktitle={Proceedings of the twenty-first International Conference on Machine learning},
  pages={114},
  year={2004}
}

@book{learning2009dataset,
  title     = {Dataset Shift in Machine Learning},
  editor    = {Joaquin Qui{\~n}onero-Candela and Masashi Sugiyama and Anton Schwaighofer and Neil D. Lawrence},
  year      = {2008},
  publisher = {The MIT Press},
  address   = {Cambridge, MA},
  isbn      = {9780262170055},
  doi       = {10.7551/mitpress/9780262170055.001.0001}
}

@inproceedings{Mesaros2018_DCASE,
    Author = "Mesaros, Annamaria and Heittola, Toni and Virtanen, Tuomas",
    title = "A multi-device dataset for urban acoustic scene classification",
    year = "2018",
    booktitle = "Proceedings of the Detection and Classification of Acoustic Scenes and Events 2018 Workshop (DCASE2018)",
    month = "November",
    pages = "9--13"
}

@inproceedings{zhang2025hybrid,
  title={Hybrid Disagreement-Diversity Active Learning for Bioacoustic Sound Event Detection},
  author={Zhang, Shiqi and Virtanen, Tuomas},
  booktitle={2025 33rd European Signal Processing Conference (EUSIPCO)},
  pages={131--135},
  year={2025},
  organization={IEEE}
}

@article{nguyen2022measure,
  title={How to measure uncertainty in uncertainty sampling for active learning},
  author={Nguyen, Vu-Linh and Shaker, Mohammad Hossein and H{\"u}llermeier, Eyke},
  journal={Machine Learning},
  volume={111},
  number={1},
  pages={89--122},
  year={2022},
  publisher={Springer}
}

@inproceedings{yang2016active,
  title={Active learning using uncertainty information},
  author={Yang, Yazhou and Loog, Marco},
  booktitle={2016 23rd International Conference on Pattern Recognition (ICPR)},
  pages={2646--2651},
  year={2016},
  organization={IEEE}
}

@article{jung2024active,
  title={Active learning of neural network potentials for rare events},
  author={Jung, Gang Seob and Choi, Jong Youl and Lee, Sangkeun Matthew},
  journal={Digital Discovery},
  volume={3},
  number={3},
  pages={514--527},
  year={2024},
  publisher={Royal Society of Chemistry}
}







\end{document}